# Dust-acoustic waves and stability in the permeating dusty plasma: II. Power-law distributions

Jingyu Gong[1], Zhipeng Liu[1,2], and Jiulin Du[1, a)]

[1]*Department of Physics, School of Science, Tianjin University, Tianjin 300072, China*

[2]*Department of Fundamental Subject, Tianjin Institute of Urban Construction, Tianjin 300384, China*

**Abstract**

The dust-acoustic waves and their stability driven by a flowing dusty plasma when it cross through a static (target) dusty plasma (the so-called permeating dusty plasma) are investigated when the components of the dusty plasma obey the power-law $q$-distributions in nonextensive statistics. The frequency, the growth rate and the stability condition of the dust-acoustic waves are derived under this physical situation, which express the effects of the nonextensivity as well as the flowing dusty plasma velocity on the dust-acoustic waves in this dusty plasma. The numerical results illustrate some new characteristics of the dust-acoustic waves, which are different from those in the permeating dusty plasma when the plasma components are the Maxwellian distribution. In addition, we show that the flowing dusty plasma velocity has a significant effect on the dust-acoustic waves in the permeating dusty plasma with the power-law $q$-distribution.

---

a) Electronic mail: jiulindu@yahoo.com.cn

## I. INTRODUCTION

As reviewed in paper I[1], the permeating (interpenetrating) dusty plasma system consists of two parts; one part is the flowing dusty plasma, the other is the target dusty plasma being in a relative static state. The flowing dusty plasma moves through the target dusty plasma when they are encounter in space. The dust-acoustic waves and their stability in the dusty plasma can be analyzed on the basis of the related statistical theory. It has been observed that the electrons, ions, and dust grains either may be the Maxwellian distribution in some physical situations near equilibrium, and they may also be the non-Maxwellian distributions in many physical situations far away from equilibrium[2-8]. If the components of dusty plasma are the non-Maxwellian distributions or the power-law $q$-distributions that can be described by nonextensive statistics, the characteristics will be different from those described in the realm governed by the traditional statistical mechanics[9-13]. Recently, the plasma with the power-law $q$-distributions have been investigated in the framework of nonextensive statistical mechanics, such as the basic characteristics of the plasma[14-23], the solar wind and space plasma[24-30], and the astrophysical plasma[31-36], etc. It has been known that the power-law $q$-distributions have represented some basic characteristics of the plasma in complex systems away from equilibrium.

In paper I[1], we have investigated the dust-acoustic waves in the permeating dusty plasma and their stability if the components of the dusty plasma obey the Maxwellian distributions in the traditional statistics. In this paper, we investigate the above characteristics of the permeating dusty plasma if the components obey the power-law $q$-distributions in nonextensive statistics. The basic theory for the dust-acoustic waves of the plasma is described in Sec.II. The characteristics of the permeating dusty plasma with the power-law $q$-distribution are investigated in Sec.III. In Sec.IV, we present the numerical investigation on these characteristics, and finally in Sec. V, we summarize the conclusions.

## II. THE BASIC THEORY FOR THE DUST-ACOUSTIC WAVES IN THE DUSTY PLASMA

Firstly, let us present a brief review on the basic equations and the dispersion function for the dust-acoustic waves in a collisionless, unbounded, unmagnetized dusty plasma, composed of electrons, ions and dusty grains. The kinetic equation for the particles can be written by the linearized Vlasov equation[36],

$$\frac{\partial f_{\alpha 1}}{\partial t} + \mathbf{v} \cdot \nabla f_{\alpha 1} + \frac{Q_\alpha}{m_\alpha} \mathbf{E}_1 \cdot \nabla_v f_{\alpha 0} = 0, \quad (1)$$

where $f_{\alpha 1}(\mathbf{r},\mathbf{v},t)$ denotes a perturbation about the equilibrium distribution $f_{\alpha 0}(\mathbf{v})$ of the particles, with $\alpha = d, i, e$ stands, respectively, for the dusty grain, the ion and the electron. $\mathbf{E}_1$ is the electric field produced by the perturbation, and $Q_\alpha$ is the charge of the component $\alpha$. Correspondingly, the linearized Poisson equation is

$$\nabla \cdot \mathbf{E}_1 = \frac{1}{\varepsilon_0} \sum_\alpha Q_\alpha \int f_{\alpha 1} d\mathbf{v}. \quad (2)$$

Because $f_{\alpha 1}(\mathbf{r},\mathbf{v},t) \propto \exp[i(\mathbf{k}\cdot\mathbf{r} - \omega t)]$ and $\mathbf{E}_1 \propto \exp[i(\mathbf{k}\cdot\mathbf{r} - \omega t)]$, by making Fourier transformation for *r* and Laplace transformation for *t* in the Eqs.(1) and (2), one obtains

$$-i\omega f_{\alpha 1} + i\mathbf{k}\cdot\mathbf{v} f_{\alpha 1} + \frac{q_\alpha}{m_\alpha} \mathbf{E}_1 \cdot \nabla_v f_{\alpha 0} = 0, \quad (3)$$

$$i\mathbf{k}\cdot\mathbf{E}_1 = \frac{1}{\varepsilon_0} \sum_\alpha q_\alpha \int f_{\alpha 1} d\mathbf{v}. \quad (4)$$

If the wave vector $\mathbf{k}$ is along *x*-axis, $v_x = u$, the oscillating frequency is $\omega_{p\alpha} = \sqrt{n_{\alpha 0} Q_\alpha^2 / \varepsilon_0 m_\alpha}$ and the normalized equilibrium distribution is $\hat{f}_{\alpha 0} = f_{\alpha 0}/n_{\alpha 0}$, according to Landau path integral[38], it is derived that the dispersion relation is

$$1 + \sum_\alpha \frac{\omega_{p\alpha}^2}{k^2} \int \frac{\partial \hat{f}_{\alpha 0}/\partial u}{\omega/k - u} du = 0. \quad (5)$$

Or, equivalently, it can be written as

$$\varepsilon(\omega,k) = 1 + \sum_{\alpha} \chi_{\alpha} = 0, \tag{6}$$

where the physical quantity

$$\chi_{\alpha} = \frac{\omega_{p\alpha}^2}{k^2} \int \frac{\partial \hat{f}_{\alpha 0}/\partial u}{\omega/k - u} \, du \tag{7}$$

is called polarizability. Based on the above theory, we can study the dust-acoustic waves and their stability in permeating dusty plasma.

## III. THE DUST-ACOUSTIC WAVES AND STABILITY IN THE PERMEATING DUSTY PLASMA WITH POWER-LAW DISTRIBUTION

If the permeating dusty plasma is not near thermal equilibrium, but is far from the equilibrium, the particles will deviate from the Maxwellian distribution, and in many physical situations they may be the kappa distribution, which can be described equivalently by the power-law $q$-distribution in nonextensive statistics[9,14-16]. The power-law velocity $q$-distribution[9,14-16] for one-dimensional case can be written by

$$\hat{f}_{\alpha 0} = \hat{f}_{\alpha}^q = \frac{A_{q\alpha}}{\sqrt{2\pi} v_{T\alpha}} \cdot \left\{ 1 - (q_{\alpha}-1) \frac{(v-v_{\alpha 0})^2}{v_{T\alpha}^2} \right\}^{\frac{1}{q_{\alpha}-1}} \tag{8}$$

with 
$$A_{q\alpha} = \sqrt{1-q_{\alpha}} \, \frac{\Gamma\left(\frac{1}{1-q_{\alpha}}\right)}{\Gamma\left(\frac{1}{1-q_{\alpha}} - \frac{1}{2}\right)} \quad \text{for } 0 < q_{\alpha} \leq 1,$$

and 
$$A_{q\alpha} = \frac{1+q_{\alpha}}{2} \sqrt{q_{\alpha}-1} \, \frac{\Gamma\left(\frac{1}{q_{\alpha}-1} + \frac{1}{2}\right)}{\Gamma\left(\frac{1}{q_{\alpha}-1}\right)} \quad \text{for } q_{\alpha} > 1.$$

This power-law velocity $q$-distribution recovers the Maxwellian velocity distribution Eq.(8) given in paper I[1] when it is in the limit $q_{\alpha} \to 1$. The $q$-parameter $q_{\alpha}$ in this distribution may be related to the temperature gradient $\nabla T_{\alpha}$ and the Coulombian potential $\varphi$ by the following relation[7,9,14-16],

$$k_B \nabla T_{\alpha} + (1-q_{\alpha}) Q_{\alpha} \nabla \varphi = 0 \tag{9}$$

where $k_B$ is the Boltzmann constant, and $Q_\alpha$ is the charge of the component $\alpha$. In the unmagnetized dust plasma, the magnitude of potential gradient $\nabla\varphi$ is equal to that of electric field $E_1$, but the sign is opposite. The electric field is determined by the linearized Poisson equation Eq.(2). Thus, the nonextensive parameter $q_\alpha \neq 1$ as well as the power-law $q$-distribution is shown to describe the characteristic of the plasma system being a nonequilibrium stationary-state when it is away from the thermal equilibrium.

Under this power-law velocity $q$-distribution, the dispersion function for the plasma is generalized[9,15,16,23] by

$$Z_{q\alpha}(\xi_\alpha) = \frac{A_{q\alpha}}{\sqrt{\pi}} \int_{-\infty}^{\infty} \frac{\left[1-(q_\alpha-1)x^2\right]^{\frac{2-q_\alpha}{q_\alpha-1}}}{x-\xi_\alpha} \, dx \qquad (10)$$

The integrand in Eq.(10) has the same singular point as that in Eq.(9) in paper I[1], i.e. $x = \xi_\alpha$ on the $x$-axis. According to Landau integral path, we can get

$$Z_{q\alpha}(\xi_\alpha) = \frac{A_{q\alpha}}{\sqrt{\pi}} \Pr \int_{-\infty}^{\infty} \frac{\left[1-(q_\alpha-1)x^2\right]^{\frac{2-q_\alpha}{q_\alpha-1}}}{x-\xi_\alpha} \, dx + iA_{q\alpha}\sqrt{\pi}\left[1-(q_\alpha-1)\xi_\alpha^2\right]^{\frac{2-q_\alpha}{q_\alpha-1}}. \qquad (11)$$

Then the dispersion relation Eq.(5) can be generalized as

$$1 + \sum_\alpha \frac{1}{k^2\lambda_{D\alpha}^2}\left\{\frac{1+q_\alpha}{2} + \xi_\alpha \frac{A_{q\alpha}}{\sqrt{\pi}} \Pr \int_{-\infty}^{\infty} \frac{\left[1-(q_\alpha-1)x^2\right]^{\frac{2-q_\alpha}{q_\alpha-1}}}{x-\xi_\alpha} \, dx + i\xi_\alpha A_{q\alpha}\sqrt{\pi}\left[1-(q_\alpha-1)\xi_\alpha^2\right]^{\frac{2-q_\alpha}{q_\alpha-1}}\right\} = 0, \qquad (12)$$

which can be written as $\varepsilon_q(\omega,k) = 0$, also called the dielectric function.

We now consider the same physical situation as that for Eqs.(14)-(15) in paper I[1]. And we still let the target plasma be static, $v_{s\alpha} = 0$ and every component of the flowing plasma be with the same velocity, i.e. $v_{f\alpha} = v_{f0}$, and we make use of the condition: $v_{Tsd} \ll v_\phi \ll v_{Tsi}, v_{Tse}$ and $v_\phi - v_{f0} \ll v_{Tsd}, v_{Tsi}, v_{Tse}$, where $v_\phi = \omega/k$ is the phase velocity of the dust-acoustic wave. In this way, the real part Eq.(16) and the imaginary part Eq.(17) of the dielectric function in paper I[1] are now generalized to the permeating dusty plasma with the power-law $q$-distribution. The real part becomes

$$\varepsilon_{qr}(\omega,k) \approx 1 + \frac{1}{k^2\lambda_{Dq}^2} - \frac{\omega_{psd}^2}{\omega^2} - \frac{3k^2 v_{Tsd}^2 \omega_{psd}^2}{(3q_{sd}-1)\omega^4}, \quad (13)$$

and the imaginary part becomes

$$\varepsilon_{qi}(\omega,k) \approx 2\sqrt{\pi}\left[A_{qsd}\frac{\omega\omega_{psd}^2}{k^3 v_{Tsd}^3}\left[1-(q_{sd}-1)\frac{\omega^2}{k^2 v_{Tsd}^2}\right]^{\frac{2-q_{sd}}{q_{sd}-1}} + A_{qsi}\frac{\omega\omega_{psi}^2}{k^3 v_{Tsi}^3} + A_{qse}\frac{\omega\omega_{pse}^2}{k^3 v_{Tse}^3}\right.$$

$$\left. + (\omega - kv_{f0})\left(A_{qfd}\frac{\omega_{pfd}^2}{k^3 v_{Tfd}^3} + A_{qfi}\frac{\omega_{pfi}^2}{k^3 v_{Tfi}^3} + A_{qfe}\frac{\omega_{pfe}^2}{k^3 v_{Tfe}^3}\right)\right], \quad (14)$$

where $\frac{1}{\lambda_{Dq}^2} = \frac{1+q_{si}}{2\lambda_{Dsi}^2} + \frac{1+q_{se}}{2\lambda_{Dse}^2} + \frac{1+q_{fd}}{2\lambda_{Dfd}^2} + \frac{1+q_{fi}}{2\lambda_{Dfi}^2} + \frac{1+q_{fe}}{2\lambda_{Dfe}^2}$. Let Eq.(13) be $\varepsilon_{qr}(\omega_{qr},k) = 0$,

we can get the dispersion function of the dust-acoustic waves,

$$\frac{\omega_{qr}^2}{k^2} = \frac{(1-3q_{sd})\omega_{psd}^2 \lambda_{Dq}^2}{\left[(1+q_{sd})(1-3q_{sd}) + 2(1-3q_{sd})k^2\lambda_{Dq}^2\right]}$$

$$\cdot\left(1 + \sqrt{1 + 3v_{Tsd}^2 \cdot \frac{\left[(1+q_{sd})(1-3q_{sd}) + 2(1-3q_{sd})k^2\lambda_{Dq}^2\right]}{(1-3q_{sd})\omega_{psd}^2 \lambda_{Dq}^2}}\right). \quad (15)$$

For the case of $v_{Tsd} \ll v_\phi \ll v_{Tsi}, v_{Tse}$, and for $q_{sd} \neq 1/3$, we obtain the frequency,

$$\omega_{qr} = \frac{\omega_{psd} k \lambda_{Dq}}{\sqrt{1+k^2\lambda_{Dq}^2}}, \quad (16)$$

where we have discarded the small fourth term on the right side of Eq.(13). Thus we can write the normalized frequency as

$$\frac{\omega_{qr}}{\omega_{psd}} = \left\{1 + \frac{1}{2k^2\lambda_{Dsi}^2}\left[(1+q_{se})\frac{n_{se0}Q_{se}^2 T_{si}}{n_{si0}Q_{si}^2 T_{se}} + (1+q_{fd})\frac{n_{fd0}Q_{fd}^2 T_{si}}{n_{si0}Q_{si}^2 T_{fd}}\right.\right.$$

$$\left.\left. + (1+q_{fi})\frac{n_{fi0}Q_{fi}^2 T_{si}}{n_{si0}Q_{si}^2 T_{fi}} + (1+q_{fe})\frac{n_{fe0}Q_{fe}^2 T_{si}}{n_{si0}Q_{si}^2 T_{fe}} + (1+q_{si})\right]\right\}^{-\frac{1}{2}}, \quad (17)$$

where the parameter $q_{sd}$ disappear, which implies that the $q$-distribution of the dusty gains in the target plasma doesn't influence the normalized frequency.

The growth rate of the dust-acoustic waves can be obtained from Eq.(13) and Eq.(14). It is derived that

$$\gamma_q = -\frac{\varepsilon_i(\omega_r, k)}{\left.\dfrac{\partial \varepsilon_r}{\partial \omega}\right|_{\omega=\omega_{qr}}}$$

$$= -\frac{\sqrt{\pi}\,\omega_{qr}}{1 + \dfrac{6}{3q_{sd}-1}\dfrac{k^2 v_{Tsd}^2}{\omega_{qr}^2}} \Bigg\{ \frac{A_{qsd}\omega_{qr}^3}{k^3 v_{Tsd}^3}\left[1-(q_{sd}-1)\frac{\omega_{qr}^2}{k^2 v_{Tsd}^2}\right]^{\frac{2-q_{sd}}{q_{sd}-1}} + \frac{A_{qsi}\omega_{qr}^3 \omega_{psi}^2}{k^3 v_{Tsi}^3 \omega_{psd}^2}$$

$$+ \frac{A_{qse}\omega_{qr}^3 \omega_{pse}^2}{k^3 v_{Tse}^3 \omega_{psd}^2} + \left(\frac{\omega_{qr}^3}{k^3} - \frac{\omega_{qr}^2}{k^2} v_{f0}\right)\left(\frac{A_{qfd}\omega_{pfd}^2}{v_{Tfd}^3 \omega_{psd}^2} + \frac{A_{qfi}\omega_{pfi}^2}{v_{Tfi}^3 \omega_{psd}^2} + \frac{A_{qfe}\omega_{pfe}^2}{v_{Tfe}^3 \omega_{psd}^2}\right)\Bigg\}. \tag{18}$$

Using $v_{Tsd}^2 = 2\omega_{psd}^2 \lambda_{Dsd}^2$ and substituting Eq.(17) into Eq.(18), we can write the normalized growth rate as

$$\frac{\gamma_q}{\omega_{psd}} = -B_q\left[C_q\left(1-\frac{v_{f0}}{v_{q\phi}}\right) + D_q\right], \tag{19}$$

with the $q$-related parameters:

$$B_q = \frac{\sqrt{\pi}}{2\sqrt{2}}\left(\frac{n_{sd0}Q_{sd}^2 T_{si}}{n_{si0}Q_{si}^2 T_{sd}}\right)^{\frac{3}{2}} k\lambda_{Dsi}\Bigg[ k^2\lambda_{Dsi}^2 + \frac{1+q_{si}}{2} + \frac{1+q_{se}}{2}\frac{n_{se0}Q_{se}^2 T_{si}}{n_{si0}Q_{si}^2 T_{se}}$$

$$+ \frac{1+q_{fd}}{2}\frac{n_{fd0}Q_{fd}^2 T_{si}}{n_{si0}Q_{si}^2 T_{fd}} + \frac{1+q_{fi}}{2}\frac{n_{fi0}Q_{fi}^2 T_{si}}{n_{si0}Q_{si}^2 T_{fi}} + \frac{1+q_{fe}}{2}\frac{n_{fe0}Q_{fe}^2 T_{si}}{n_{si0}Q_{si}^2 T_{fe}}\Bigg]^{-2}$$

$$\Bigg[1 + \frac{12}{3q_{sd}-1}\frac{n_{si0}Q_{si0}^2 T_{sd}}{n_{sd0}Q_{sd0}^2 T_{si}}\cdot\left(k^2\lambda_{Dsi}^2 + \frac{1+q_{si}}{2} + \frac{1+q_{se}}{2}\frac{n_{se0}Q_{se}^2 T_{si}}{n_{si0}Q_{si}^2 T_{se}}\right.$$

$$\left.+ \frac{1+q_{fd}}{2}\frac{n_{fd0}Q_{fd}^2 T_{si}}{n_{si0}Q_{si}^2 T_{fd}} + \frac{1+q_{fi}}{2}\frac{n_{fi0}Q_{fi}^2 T_{si}}{n_{si0}Q_{si}^2 T_{fi}} + \frac{1+q_{fe}}{2}\frac{n_{fe0}Q_{fe}^2 T_{si}}{n_{si0}Q_{si}^2 T_{fe}}\right)\Bigg]^{-1}, \tag{20}$$

$$C_q = \frac{A_{qfd} n_{fd0} Q_{fd0}^2 m_{fd}^{\frac{1}{2}} T_{sd}^{\frac{3}{2}}}{n_{sd0} Q_{sd0}^2 m_{sd}^{\frac{1}{2}} T_{fd}^{\frac{3}{2}}} + \frac{A_{qfi} n_{fi0} Q_{fi0}^2 m_{fi}^{\frac{1}{2}} T_{sd}^{\frac{3}{2}}}{n_{sd0} Q_{sd0}^2 m_{sd}^{\frac{1}{2}} T_{fi}^{\frac{3}{2}}} + \frac{A_{qfe} n_{fe0} Q_{fe0}^2 m_{fe}^{\frac{1}{2}} T_{sd}^{\frac{3}{2}}}{n_{sd0} Q_{sd0}^2 m_{sd}^{\frac{1}{2}} T_{fe}^{\frac{3}{2}}}, \tag{21}$$

and

$$D_q = A_{qsd}\Bigg\{1 - \frac{(q_{sd}-1)}{2}\frac{n_{sd0}Q_{sd0}^2 T_{si}}{n_{si0}Q_{si0}^2 T_{sd}}\Bigg[k^2\lambda_{Dsi}^2 + \frac{1+q_{si}}{2} + \frac{1+q_{se}}{2}\frac{n_{se0}Q_{se}^2 T_{si}}{n_{si0}Q_{si}^2 T_{se}}$$

$$+ \frac{1+q_{fd}}{2}\frac{n_{fd0}Q_{fd}^2 T_{si}}{n_{si0}Q_{si}^2 T_{fd}} + \frac{1+q_{fi}}{2}\frac{n_{fi0}Q_{fi}^2 T_{si}}{n_{si0}Q_{si}^2 T_{fi}} + \frac{1+q_{fe}}{2}\frac{n_{fe0}Q_{fe}^2 T_{si}}{n_{si0}Q_{si}^2 T_{fe}}\Bigg]^{-1}\Bigg\}^{\frac{2-q_{sd}}{q_{sd}-1}}$$

$$+\frac{A_{qsi}n_{si0}Q_{si0}^2 m_{si}^{\frac{1}{2}} T_{sd}^{\frac{3}{2}}}{n_{sd0}Q_{sd0}^2 m_{sd}^{\frac{1}{2}} T_{si}^{\frac{3}{2}}}+\frac{A_{qse}n_{se0}Q_{se0}^2 m_{se}^{\frac{1}{2}} T_{sd}^{\frac{3}{2}}}{n_{sd0}Q_{sd0}^2 m_{sd}^{\frac{1}{2}} T_{se}^{\frac{3}{2}}}. \tag{22}$$

From Eq.(19) we find that the instability condition of the dust-acoustic waves in the permeating plasma with the power-law $q$-distributions is given by

$$v_{f0} > v_{q\phi}\left(1+\frac{D_q}{C_q}\right). \tag{23}$$

As compared with Eq.(27) in paper I[1], the phase velocity $v_{q\phi}$, the parameters $C_q$ and $D_q$ now depend on three $q$-parameters for the three components with the power-law q-distributions. Eq.(27) in paper I[1] can be recovered by Eq.(23) only when the three $q$-parameters are set to unity.

If the dust plasma system is without the permeating phenomenon, Eq.(18) becomes

$$\gamma_q = -\frac{\sqrt{\pi}\omega_{qr}}{1+\frac{6}{3q_d-1}\frac{k^2 v_{Td}^2}{\omega_{qr}^2}}\left\{\frac{A_{qd}\omega_{qr}^3}{k^3 v_{Td}^3}\left[1-(q_d-1)\frac{\omega_{qr}^2}{k^2 v_{Td}^2}\right]^{\frac{2-q_d}{q_d-1}}+\frac{A_{qi}\omega_{qr}^3 \omega_{pi}^2}{k^3 v_{Tsi}^3 \omega_{pd}^2}+\frac{A_{qe}\omega_{qr}^3 \omega_{pe}^2}{k^3 v_{Tse}^3 \omega_{pd}^2}\right\}, \tag{24}$$

where we have used the condition, $v_{Td} \ll v_\phi \ll v_{Ti}, v_{Te}$.

Again we consider the physical situation, i.e. if the flowing velocities $v_{f\alpha}$ are different from each other for each component $\alpha$ of the flowing dusty plasma, we have $v_\phi \ll v_{Tsd}+v_{fd}, v_{Tsi}+v_{fi}, v_{Tse}+v_{fe}$, and the growth rate Eq.(18) is revised as

$$\gamma_q = -\frac{\sqrt{\pi}\omega_{qr}}{1+\frac{6}{3q_{sd}-1}\frac{k^2 v_{Tsd}^2}{\omega_{qr}^2}}\left\{\frac{A_{qsd}\omega_{qr}^3}{k^3 v_{Tsd}^3}\left[1-(q_{sd}-1)\frac{\omega_{qr}^2}{k^2 v_{Tsd}^2}\right]^{\frac{2-q_{sd}}{q_{sd}-1}}+\frac{A_{qsi}\omega_{qr}^3 \omega_{psi}^2}{k^3 v_{Tsi}^3 \omega_{psd}^2}\right.$$

$$\left.+\frac{A_{qse}\omega_{qr}^3 \omega_{pse}^2}{k^3 v_{Tse}^3 \omega_{psd}^2}+\sum_{\alpha=d,i,e}\left(\frac{\omega_{qr}^3}{k^3}-\frac{\omega_{qr}^2}{k^2}v_{f\alpha 0}\right)\frac{A_{qf\alpha}\omega_{pf\alpha}^2}{v_{Tf\alpha}^3 \omega_{psd}^2}\right\}. \tag{25}$$

Thus Eq.(19) becomes

$$\frac{\gamma_q}{\omega_{psd}} = -B_q \left[ D_q + \sum_{\alpha=d,i,e} C_{q\alpha} \left(1 - \frac{v_{f\alpha 0}}{v_{q\phi}}\right) \right], \quad (26)$$

where

$$C_{q\alpha} = A_{qf\alpha} \frac{n_{f\alpha 0} Q_{f\alpha 0}^2 m_{f\alpha}^{\frac{1}{2}} T_{sd}^{\frac{3}{2}}}{n_{sd0} Q_{sd0}^2 m_{sd}^{\frac{1}{2}} T_{f\alpha}^{\frac{3}{2}}}. \quad (27)$$

Again using $m_d \gg m_i, m_e$, then $C_{qi}$ and $C_{qe}$ are small as compared with $C_{qd}$ so that they can be discarded in Eq.(26). The instability condition of the dust-acoustic waves now becomes

$$v_{fd} > v_{q\phi} \left(1 + \frac{D_q}{C_{qd}}\right). \quad (28)$$

It is clear that the parameters $q_{sd}$, $q_{si}$, $q_{se}$, $q_{fd}$ may play important roles in the instability condition of the waves.

## IV. NUMERICAL CALCULATIONS

In order to illustrate our theoretical results more clearly, we now perform numerical calculations for the frequency and the growth rate of the dust-acoustic wave, and the critical flowing velocity for the instability condition in the permeating dusty plasma.

For the numerical calculations on the permeating dusty plasma with the power law $q$-distribution, we consider the same physical case as those in paper I[1], in which the dusty plasma matches the conditions for the particle's number densities, masses, and charges, i.e. the particle's number density of each component at the equilibrium satisfies $n_{sd0} = n_{fd0} = n_{d0}$, $n_{si0} = n_{fi0} = n_{i0}$, and $n_{se0} = n_{fe0} = n_{e0}$; the particle's mass of each component satisfies $m_{sd} = m_{fd} = m_d$, $m_{si} = m_{fi} = m_i$ and $m_{se} = m_{fe} = m_e$, and the number density rate between the ion and electron is $n_{i0}/n_{e0} = \delta$ at the equilibrium. Typical density parameters of dust-laden plasma in interstellar clouds[39-41] are $n_e \sim 10^{-3} \text{cm}^{-3}$ and $n_d \sim 10^{-7} \text{cm}^{-3}$. The dust grains in interstellar clouds are dielectric (ices, silicates, etc) and metallic (graphite, magnetite, amorphous carbons, etc), and the radii of the dust grains are approximately $r_d \simeq 0.2 \mu m$ for the ices. The mass rate between the electron, ion and dust grain still may be $m_e : m_i : m_d \approx 1 : 10^4 : 10^9$. The charge of the electron, ion and dust grain at the equilibrium are $Q_{e0} = Q_{i0} = e$, and $Q_{d0} = (\delta - 1) \times 10^4 e$ based on the quasineutral condition. In addition, we assume the $q$-parameters for the dust grains, the ions, and the electrons satisfy $q_{sd} = q_{fd} = q_d$, $q_{si} = q_{fi} = q_i$, and $q_{se} = q_{fe} = q_e$. Under the conditions of $T_{sd} \ll T_{si} \ll T_{se}$ and $T_{fd} \ll T_{fi} \ll T_{fe}$, ignoring the infinitesimal quantities in the related equations in Eqs.(17), (19) and (23), we can write the normalized frequency as

$$\frac{\omega}{\omega_{psd}} = \frac{k\lambda_{Dsi}}{\sqrt{k^2\lambda_{Dsi}^2 + \frac{1+q_i}{2}\left(1+\frac{T_{si}}{T_{fi}}\right) + \frac{1+q_d}{2}\frac{(\delta-1)^2 T_{si}}{\delta T_{fd}} \times 10^4}}, \quad (29)$$

the growth rate of the waves as

$$\frac{\gamma}{\omega_{psd}} = -\frac{\frac{\sqrt{\pi}k\lambda_{Dsi}}{2\sqrt{2}}\left(\frac{(\delta-1)^2 T_{si}}{\delta T_{sd}} \times 10^4\right)^{\frac{3}{2}}}{\left(k^2\lambda_{Dsi}^2 + \frac{1+q_i}{2}\left(1+\frac{T_{si}}{T_{fi}}\right) + \frac{1+q_d}{2}\frac{(\delta-1)^2 T_{si}}{\delta T_{fd}} \times 10^4\right)^2}$$

$$\cdot \left\{ A_{qd} \left[1 - \frac{(q_d - 1)(\delta - 1)^2 T_{si} \times 10^4}{2\delta T_{sd}\left(k^2\lambda_{Dsi}^2 + \frac{1+q_i}{2}\left(1+\frac{T_{si}}{T_{fi}}\right) + \frac{1+q_d}{2}\frac{(\delta-1)^2 T_{si}}{\delta T_{fd}} \times 10^4\right)}\right]^{\frac{2-q_d}{q_d-1}} \right.$$

$$\left. + \left(1 - \frac{v_{f0}}{v_\varphi}\right) A_{qd} \left(\frac{T_{sd}}{T_{fd}}\right)^{\frac{3}{2}} \right\}, \quad (30)$$

and the instability condition for the dust-acoustic waves as

$$v_{f0}/v_{q\varphi} > 1 + D_q/C_q, \quad (31)$$

where the critical value can be expressed by

$$\frac{D_q}{C_q} = \left[1 - \frac{(q_d-1)(\delta-1)^2 T_{si} \times 10^4}{2\delta T_{sd}\left(k^2\lambda_{Dsi}^2 + \frac{1+q_i}{2}\left(1+\frac{T_{si}}{T_{fi}}\right) + \frac{1+q_d}{2}\frac{(\delta-1)^2 T_{si}}{\delta T_{fd}} \times 10^4\right)}\right]^{\frac{2-q_d}{q_d-1}} \cdot \left(\frac{T_{sd}}{T_{fd}}\right)^{-\frac{3}{2}}. \quad (32)$$

Based on these equations, the numerical results are shown in Fig.1-10. In these figures, we illustrate the nonextensive parameters $q_\alpha$ of the dusty plasma and their influences on the normalized frequency, the growth rate of the dust-acoustic wave and the instability condition for the waves by taking $\delta=1.002$, $T_{sd}=10\mathrm{K}$, $T_{fd}=10^2\mathrm{K}$, $T_{si}=10^3\mathrm{K}$ and $T_{fi}=10^4\mathrm{K}$.

Fig.1 and Fig.2 illustrate the numerical results of the generalized normalized frequency, Eq.(29), for the dust-acoustic waves as a function of the normalized waves number $k\lambda_{Dsi}$, which is plotted with different values of the nonextensive parameters $q_d$ for the dust grains and $q_i$ for the ions. They show that the generalized normalized frequency is not very significantly dependent on the nonextensive parameters $q_d$ and $q_i$. The curves for $q_d=q_i=1$ are the normalized frequency for the Maxwellian distribution (the solid lines).

Fig.3 and Fig.4 illustrate the numerical results of the instability critical value of the normalized flowing plasma velocity based on Eq.(32) for the dust-acoustic waves. The instability critical value is plotted as a function of the normalized wave number $k\lambda_{Dsi}$ with different values of the nonextensive parameters $q_d$ for the dust grains and $q_i$ for the ions. The cures for $q_d=q_i=1$ are the cases for the Maxwellian distribution (the solid lines). We show that the critical value of the flowing plasma velocity is significantly dependent on the nonextensive parameters $q_d$ and $q_i$ only when the normalized wave number $k\lambda_{Dsi}$ is small.

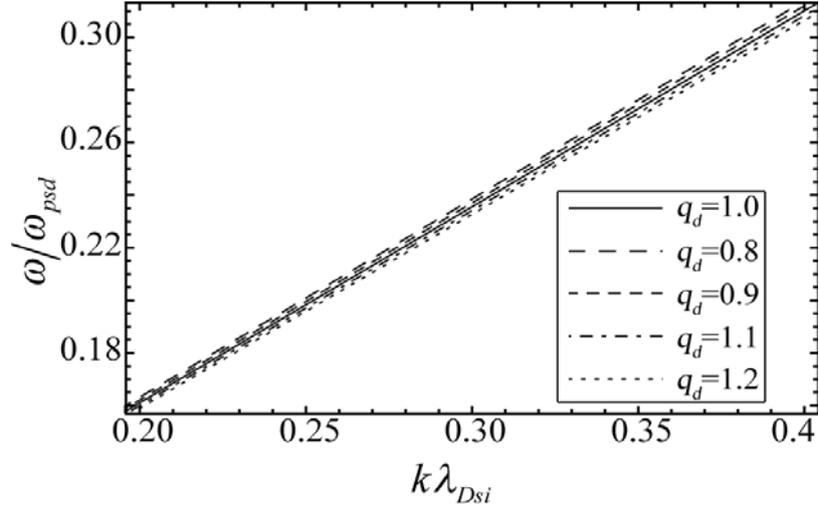

Fig.1. The normalized frequency $\omega/\omega_{psd}$ is plotted as a function of the normalized wave number $k\lambda_{Dsi}$ for different values of $q_d$ at $q_i=1$. The curves are drawn for $q_d=0.8$ (long-dashed line), $q_d=0.9$ (dashed line), $q_d=1$ (solid line), $q_d=1.1$ (dot-dashed line) and $q_d=1.2$ (dotted line).

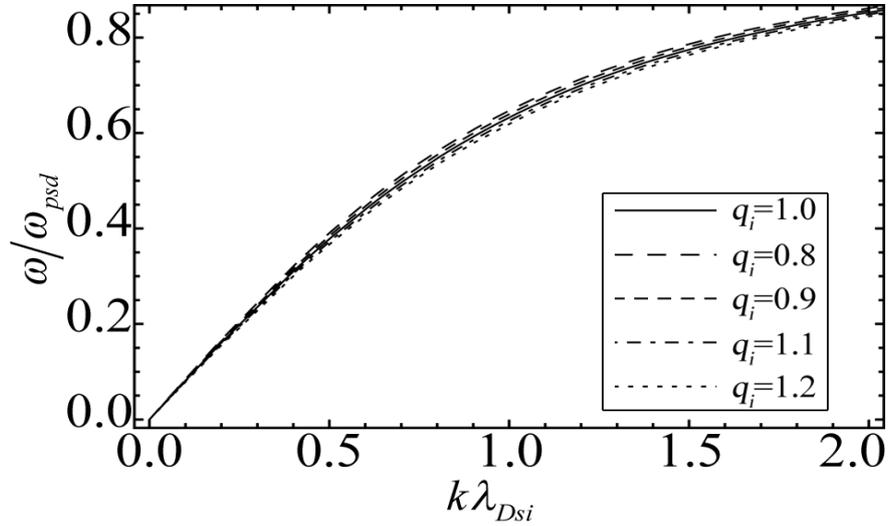

Fig.2. The normalized frequency $\omega/\omega_{psd}$ is plotted as a function of the normalized wave number $k\lambda_{Dsi}$ for different values of $q_i$ at $q_d=1$. The curves are drawn for $q_i=0.8$ (long-dashed line), $q_i=0.9$ (dashed line), $q_i=1$ (solid line), $q_i=1.1$ (dot-dashed line) and $q_i=1.2$ (dotted line).

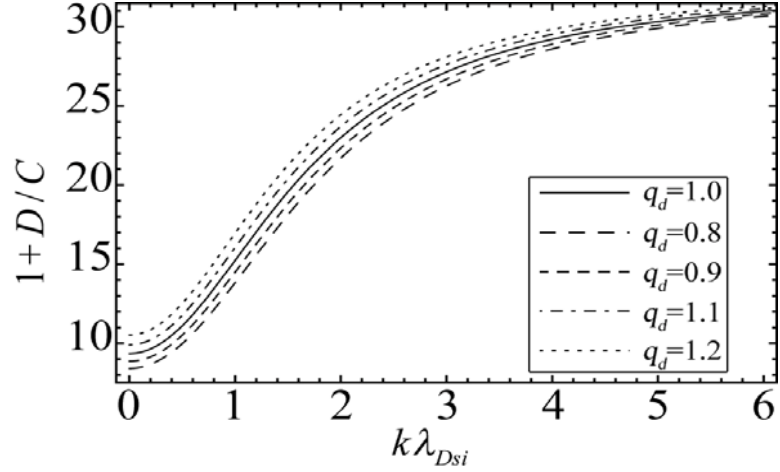

Fig.3. The instability critical values of the normalized flowing plasma velocity $v_{f0}/v_\phi$ for the dust-acoustic waves is plotted as a function of the normalized wave number $k\lambda_{Dsi}$ for different values of $q_d$ at $q_i = 1$. The curves are drawn for $q_d = 0.8$ (long-dashed line), $q_d = 0.9$ (dashed line), $q_d = 1$ (solid line), $q_d = 1.1$ (dot-dashed line) and $q_d = 1.2$ (dotted line).

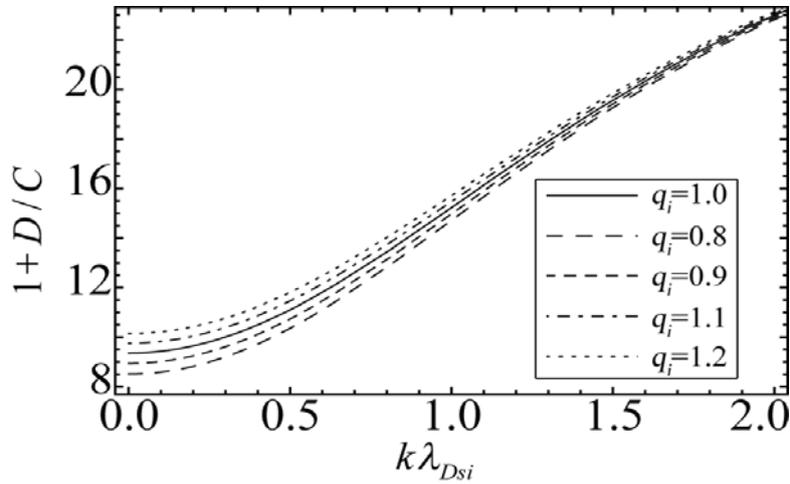

Fig.4. The instability critical values of the normalized flowing plasma velocity $v_{f0}/v_\phi$ for the dust-acoustic waves is plotted as a function of the normalized wave number $k\lambda_{Dsi}$ for different values of $q_i$ at $q_d = 1$. The curves are drawn for $q_i = 0.8$ (long-dashed line), $q_i = 0.9$ (dashed line), $q_i = 1$ (solid line), $q_i = 1.1$ (dot-dashed line) and $q_i = 1.2$ (dotted line).

Fig.5-8 are the numerical results of the growth rate $\gamma/\omega_{psd}$ as a function of $k\lambda_{Dsi}$ of the dust-acoustic waves based on Eq.(30) for the different nonextensive parmeters when we let the rate between the velocity of the flowing dusty plasma and the phase velocity be $v_{f0}/v_\phi = 40$, where $q_\alpha=1$ is the case for the Maxwellian distribution. Under this condition, the growth rate $\gamma/\omega_{psd}$ is always positive and thus the dust-acoustic waves are unstable. In Fig.5, the growth rate $\gamma/\omega_{psd}$ is plotted for the different values of $q_d$ when we take $q_i=1$, which shows that the nonextensive effects generally lead to the growth rate to become decrease. In Fig.6, the growth rate $\gamma/\omega_{psd}$ is for different values of $q_i$ when we take $q_d=1$, which shows that the nonextensive effects will lead the growth rate to become decrease if $q_i>1$ and to become increase if $q_i<1$. In Fig.7, the growth rate $\gamma/\omega_{psd}$ is for the different values of $q_{sd}$ when we take $q_i=1$, $q_{fd}=1$, which shows that the nonextensive effects will lead the growth rate to become decrease if $q_{sd}>1$ and to become increase if $q_{sd}<1$. In Fig.8, the growth rate $\gamma/\omega_{psd}$ is for the different values of $q_{fd}$ when we take $q_i=1$, $q_{sd}=1$, which shows that the nonextensive effects will lead the growth rate to become decrease if $q_{fd}<1$ and to become increase if $q_{fd}>1$.

Fig.9 and Fig.10 are also the numerical results of the growth rate $\gamma/\omega_{psd}$ as a function of $k\lambda_{Dsi}$ based on Eq.(30) for the different values of $q_d$ and $q_i$ when we let the rate between the velocity of the flowing dust plasma and the phase velocity be $v_{f0}/v_\phi =15$. Under this condition, the growth rate $\gamma/\omega_{psd}$ is positive and thus the dust-acoustic waves are unstable if the wave number is small; but the growth rate $\gamma/\omega_{psd}$ is negative and thus the dust-acoustic waves are stable if the wave number is large. In Fig.9, the growth rate $\gamma/\omega_{psd}$ is for the different values of $q_d$ when we take $q_i = 1$, which shows that the nonextensive effects will lead the growth rate to become decrease if $q_d > 1$ and to become increase if $q_d < 1$. In Fig.10, the growth rate $\gamma/\omega_{psd}$ is for the different values of $q_i$ when we take $q_d = 1$, which shows that the nonextensive effects have the influence on the growth rate only when the wave number $k\lambda_{Dsi}$ is small and lead the growth rate to become decrease if $q_i > 1$ and to become increase if $q_i < 1$.

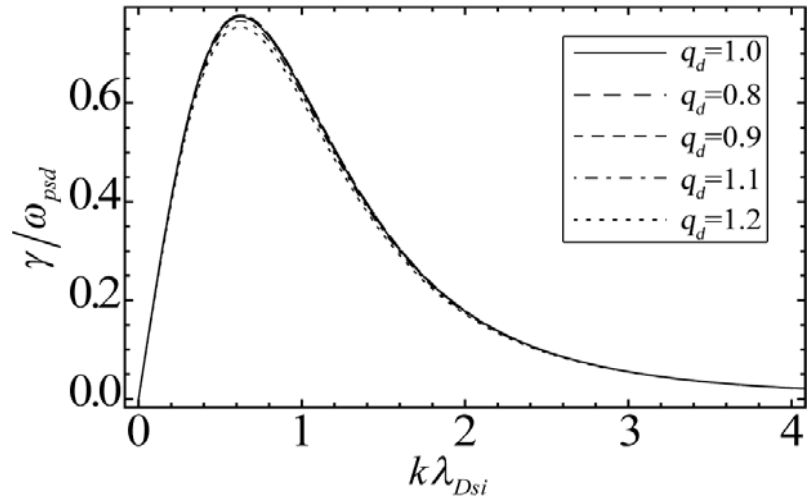

Fig.5. The growth rate of the dust-acoustic waves $\gamma/\omega_{psd}$ is plotted as a function of the normalized wave number $k\lambda_{Dsi}$ for different values of $q_d$ at $q_i = 1$ and $v_{f0}/v_\phi = 40$. The curves are drawn for $q_d = 0.8$ (long-dashed line), $q_d = 0.9$ (dashed line), $q_d = 1$ (solid line), $q_d = 1.1$ (dot-dashed line) and $q_d = 1.2$ (dotted line).

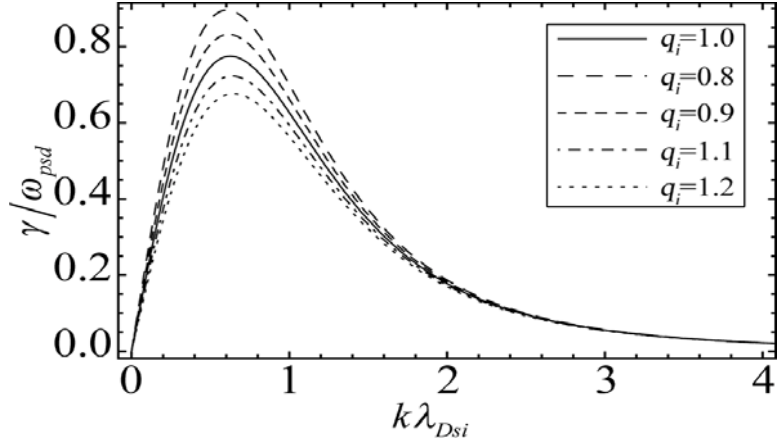

Fig.6. The growth rate of the dust-acoustic waves $\gamma/\omega_{psd}$ is plotted as a function of the normalized wave number $k\lambda_{Dsi}$ for different values of $q_i$ at $q_d = 1$, $v_{f0}/v_\phi = 40$. The curves are drawn for $q_i = 0.8$ (long-dashed line), $q_i = 0.9$ (dashed line), $q_i = 1$ (solid line), $q_i = 1.1$ (dot-dashed line) and $q_i = 1.2$ (dotted line).

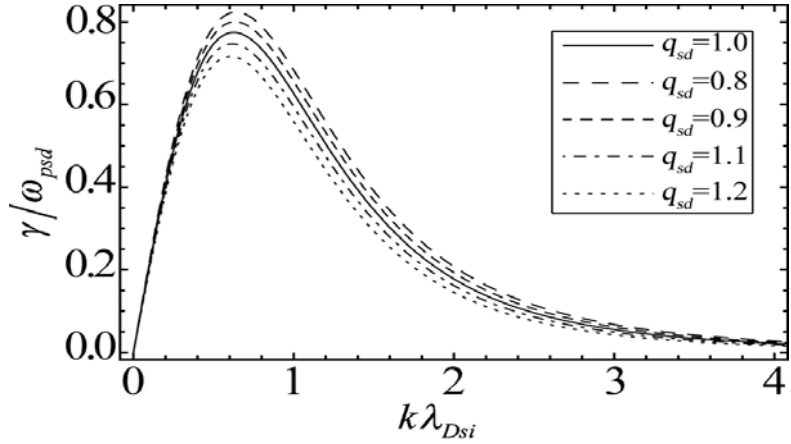

Fig.7. The growth rate of the dust-acoustic waves $\gamma/\omega_{psd}$ is plotted as a function of the normalized wave number $k\lambda_{Dsi}$ for different values of $q_{sd}$ at $q_i = 1$, $q_{fd} = 1$, $v_{f0}/v_\phi = 40$. The curves are drawn for $q_{sd} = 0.2$ (long-dashed line), $q_{sd} = 0.6$ (dashed line), $q_{sd} = 1$ (solid line), $q_{sd} = 1.4$ (dot-dashed line) and $q_{sd} = 1.8$ (dotted line).

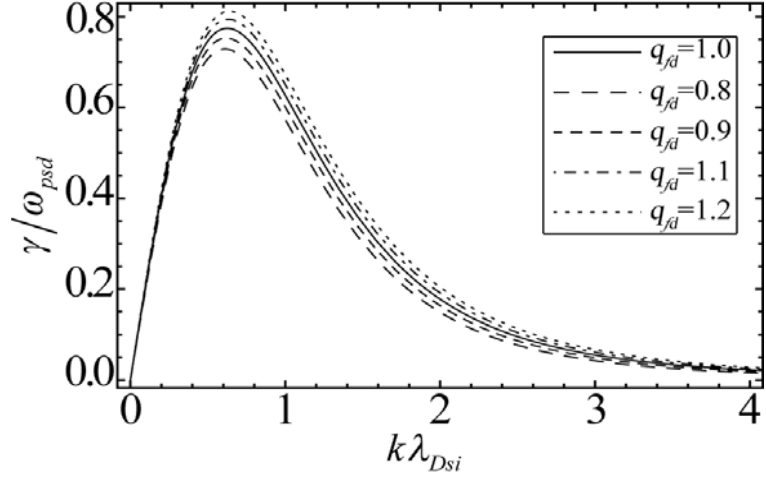

Fig.8. The growth rate of the dust-acoustic waves $\gamma/\omega_{psd}$ is plotted as a function of the normalized wave number $k\lambda_{Dsi}$ for different values of $q_{fd}$ at $q_i=1$, $q_{sd}=1$, $v_{f0}/v_\phi = 40$. The curves are drawn for $q_{fd}=0.8$(long-dashed line), $q_{fd}=0.9$ (dashed line), $q_{fd}=1$(solid line), $q_{fd}=1.1$(dot-dashed line) and $q_{fd}=1.2$(dotted line).

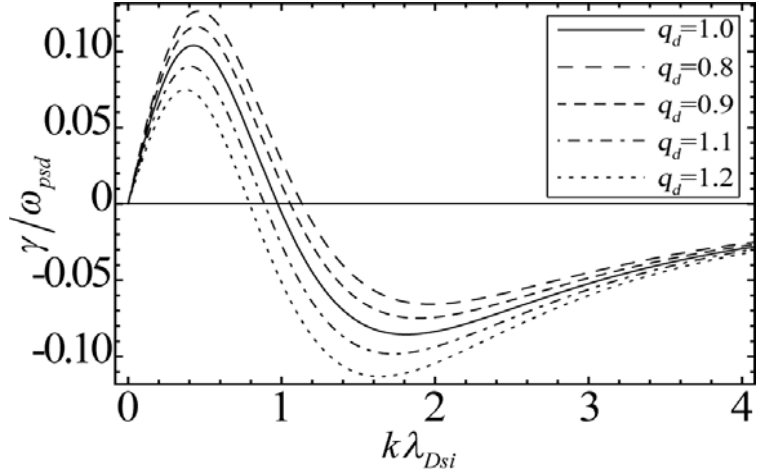

Fig.9. The growth rate of the dust-acoustic waves $\gamma/\omega_{psd}$ is plotted as a function of the normalized wave number $k\lambda_{Dsi}$ for different values of $q_d$ at $q_i=1$ and $v_{f0}/v_\phi = 15$. The curves are drawn for $q_d=0.8$(long-dashed line), $q_d=0.9$ (dashed line), $q_d=1$(solid line), $q_d=1.1$(dot-dashed line) and $q_d=1.2$(dotted line).

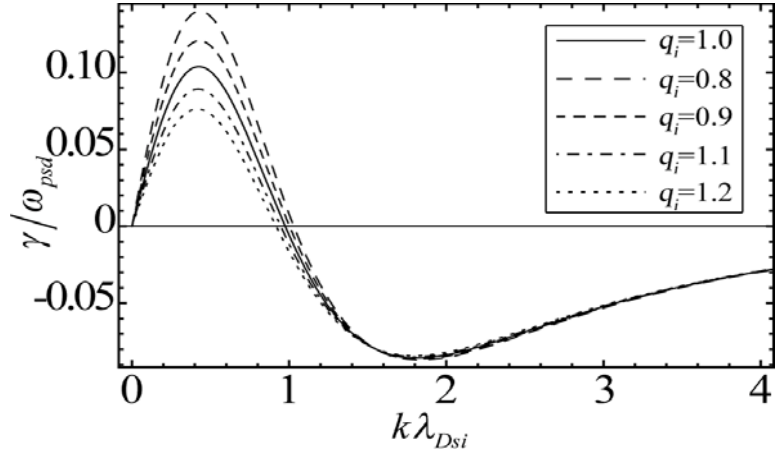

Fig.10. The growth rate of the dust-acoustic waves $\gamma/\omega_{psd}$ is plotted as a function of the normalized waves number $k\lambda_{Dsi}$ for different values of $q_i$ at $q_d = 1$ and $v_{f0}/v_\phi = 15$. The curves are drawn for $q_i = 0.8$(long-dashed line), $q_i = 0.9$ (dashed line), $q_i = 1$ (solid line), $q_i = 1.1$(dot-dashed line) and $q_i = 1.2$(dotted line).

We state that Figs.5-8 and Figs.9-10 are investigated respectively for the two rate between the velocity of the flowing dusty plasma and the phase velocity, i.e. $v_{f0}/v_\phi = 40$ and $v_{f0}/v_\phi = 15$. By comparing the results in the two cases, we find that the difference in the flowing dusty plasma velocity has a quite significant effect on the growth rate and the stability of the dust-acoustic waves in the permeating dusty plasma with the power-law $q$-distributions.

## V. CONCLUSIONS

In conclusion, we have investigated the frequency, the growth rate and the stability of the dust-acoustic waves in the permeating dusty plasma with the power-law $q$-distributions. The frequency of the dust-acoustic waves is derived by Eq.(17), the growth rate is expressed by Eq.(19), and the stability condition is obtained as Eq.(23) when the flowing velocities $v_{f\alpha}$ are considered the same for each component $\alpha$ of the flowing dusty plasma. When the flowing velocities $v_{f\alpha}$ are different from each other for each component $\alpha$, the growth rate and the instability condition are revised as Eq.(26) and Eq.(28). These equations have expressed the effects of nonextensivity in the permeating dusty plasma as well as the flowing velocity of the flowing dusty plasma on the dust-acoustic waves if the dusty plasma obeys the power-law $q$-distribution in nonextensive statistics, which is shown to have different characteristics of the dust-acoustic waves from those when the plasma is assumed to obey the traditional statistics with the Maxwellian distribution[1].

We further perform the numerical investigations on the above characteristics of the dusty plasma. The results illustrate fully the nonextensive effects ($q_\alpha$ different from unity) of each component of the permeating dusty plasma on the frequency and the growth rate of the dust-acoustic waves and their stability condition. And we do those for two different velocities of the flowing dusty plasma, and we find that the difference in the flowing dusty plasma velocity has a quite significant effect on the characteristics of the dust-acoustic waves in the permeating dusty plasma with the power-law $q$-distributions.

## ACKNOWLEDGEMENTS

This work is supported by the National Natural Science Foundation of China under Grant No.10675088 and No.11175128.